\documentclass[letterpaper]{article}
\usepackage{isea}
\usepackage[pdftex]{graphicx}
\usepackage{times}
\usepackage{helvet}
\usepackage{courier}
\usepackage[numbers]{natbib}

\title{We Are The Clouds: \\Blending Interaction and Participation in Urban Media Art}

\author{Varvara Guljajeva\textsuperscript{1,*,‡}, Mar Canet Sola\textsuperscript{2,*,‡}\\
\textsuperscript{1}The Hong Kong University of Science and Technology, Guangzhou, China, \\ \textsuperscript{2}Baltic Film, Media, and Arts School, Tallinn University\\
\\ \textsuperscript{*}These authors contributed equally to this work.\\ \textsuperscript{‡}Corresponding authors: varvarag@ust.hk, mar.canet@tlu.ee
}

\begin{document} 
 \maketitle

\begin{figure*}[h]
\includegraphics[width=\textwidth]{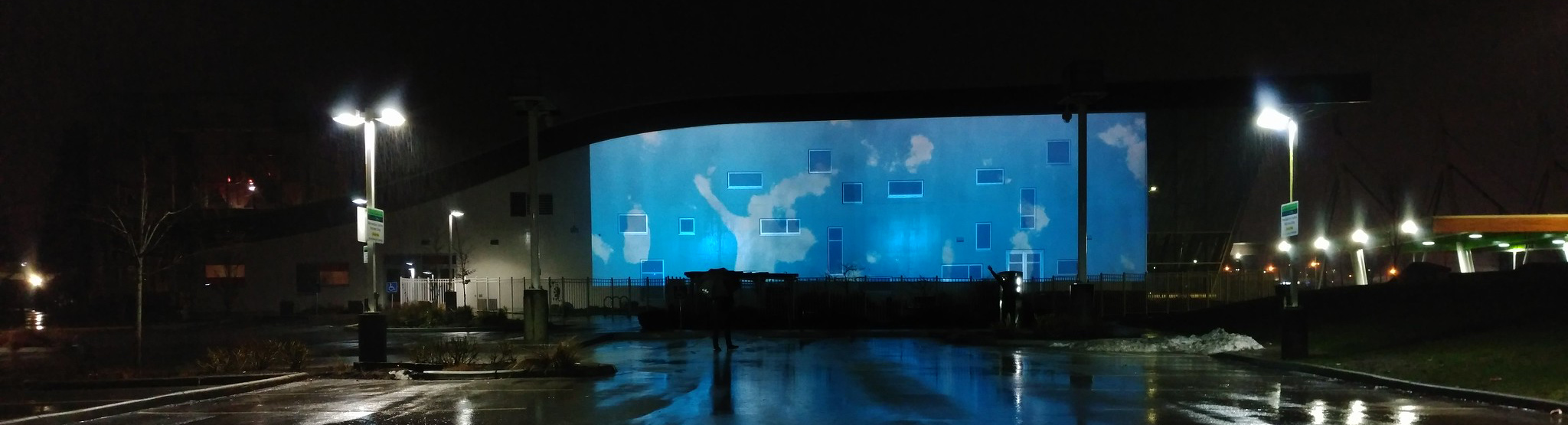}
\caption{General view of the urban media art installation \emph{We are the Clouds} on UrbanSceen at Chuck Bailey Recreation Centre in Surrey, Vancouver. \copyright Varvara \& Mar}
\label{fig:general}
\end{figure*}

\begin{abstract}
Since the early 2000s, cultural institutions have been instrumental in reshaping public spaces, fostering community engagement, and nurturing artistic innovation. Central to these initiatives are audience interaction and participation concepts, yet their definitions and applications in urban media art remain nebulous. This article endeavours to demystify these terms, examining the distinct characteristics and intersections of interactive and participatory art within urban contexts. A particular emphasis is placed on artworks that harmonise both elements, exploring the motivations and outcomes of this synthesis. The case study of \emph{We Are The Clouds} serves as a focal point, exemplifying how strategic integration of interaction and participation can enhance community connection and reinvigorate public spaces. Through this analysis, the paper underscores the transformative power of urban media artworks in redefining neighbourhood experiences, empowering local voices, and revitalising the essence of public realms.

\end{abstract}

\section{Keywords}

\section{Introduction}

According to Marc Auge, places of transition, such as airports and shopping malls, are non-places. Non-places are spaces that lack cultural identity and belonging. \cite{auge2020non} Today’s situation of public media excess Annette Weintraub describes as the paradox of overload and absence where we are overloaded with content but missing the place's cultural identity. \cite{weintraub2015overload} Several authors emphasise there is a tendency for the disappearance of public spaces and their transformation into privatised spaces with control and consumption-oriented agenda. \cite{weintraub2015overload,Stubbs2012Disco}

Media facades, which started to gain popularity in the 1990s with iconic installations such as the tricolor LED ticker in 1995 at Morgan Stanley’s in Times Square \cite{MAB2023MorganStanley}, were adopted by cultural institutions a decade later as a means to counteract the privatisation of public spaces and safeguard the cultural identity of the local communities. Notable examples include Kunsthaus Graz in 2003, Ars Electronica in Linz in 2009, MediaLab Prado in Madrid in 2010, FACT Liverpool in 2010, UrbanScreen in Surrey in 2010.  These organisations have launched urban screen programs that not only engage with the local social fabric but also provide artists with a platform to experiment with new forms and concepts. As Dave Colangelo describes in his book The Building as Screen: "Space, instead of being fixed, is co-produced in the complex interplay of social and technical actors, groups and individuals, matter and memory." \cite{colangelo2019building} In a similar vein, Scott McQuire coined the earlier term ‘relational space’ in his book The Media City. \cite{mcquire2008media}

At the heart of these initiatives is the concept of creating relationships through art, particularly participatory and interactive art. These forms of art have become pivotal in curating urban media art, offering unique ways to engage local communities and explore new relational art forms. In the following sections, we will analyse interactive and participatory urban media artworks, drawing on Varvara Guljajeva's doctoral dissertation to systematically dissect and highlight the differences between audience participation and interaction. \cite{guljajevainteraction} This examination aims to shed light on how urban media art is redefining public spaces and community engagement in the contemporary urban landscape.

\section{Distinguishing Audience Interaction from Participation in Urban Media Art}

Although in contemporary art, interaction and participation are often interchanged, yet they encompass fundamentally different aspects of audience engagement. \cite{paul2011new,guljajevainteraction} As Christiane Paul highlights, media art history is often ignored by contemporary art researchers, as she writes: “one could argue that the term ‘relational aesthetics’ itself – in its reference to the relational database, which was formalised in the 1960s and has become a defining cultural form is deeply rooted in digital technologies.” \cite{paul2011new} Bourriaud strives to find new approaches to open-ended, participatory art that avoid ‘to take shelter behind Sixties art history’ \cite{bourriaud_relational_2002}.” \cite{paul2011new}

The term 'interaction', employed in diverse fields like psychology, human-computer interaction, and social studies, acquires a distinct meaning in the context of media art. It primarily concerns how the artwork responds to and engages the audience, fostering a unique participant experience and level of engagement. \cite{edmonds2018art} To categorise urban media artworks as either interactive or participatory, we draw upon the criteria outlined in Guljajeva’s dissertation. She delineates that while both interactive and participatory art aim to forge new experiences and relationships, the key difference lies in the role of technology and the immediacy of response. Interactive art is mediated by technology and elicits an immediate reaction to audience input, enabling spontaneous engagement. In contrast, participatory art is often not mediated by technology, unfolds over a set period and often lacks an immediate response. \cite{guljajevainteraction}

This chapter further explores these distinctions through the analysis of three exemplary interactive urban media artworks: \emph{Body Movies} by Rafael Lozano-Hemmer, \emph{Blinkenlight} by the Computer Chaos Club (CCC), and \emph{LummoBlocks} by the Lummo collective; and three participatory ones: \emph{Asalto} by Daniel Canogar, \emph{\#futurefridays} by Hannah Rosenfeld, and \emph{Digital Calligraffiti} by Michael Ang and Hamza Abu Ayyash.

\begin{figure*}[h]
\includegraphics[width=\textwidth]{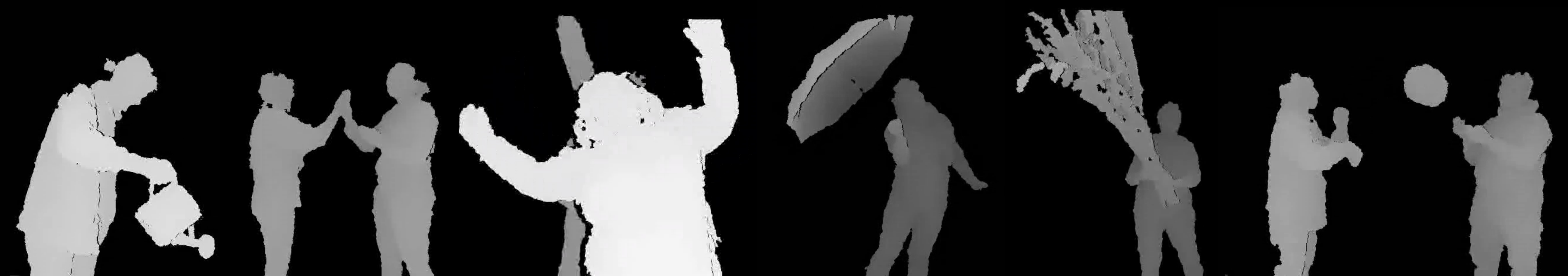}
\caption{An example of real-time rendering from silhouette video to cloud shape. \copyright Varvara \& Mar.}
\label{fig:participants_videos}
\end{figure*}

Starting with \emph{Body Movies}, it was one of the first multi-user interactive large-scale urban projections premiered in Rotterdam in 2001. The artist used the latest computer vision technology to track the audience’s bodies and projected them as shadows that revealed hidden photos of passers-by. The audience was invited to interact in real-time by performing different actions in front of the light source. The size of the audience's silhouettes depended on their distance from the light source on the ground ranging from 2 to 30m. This interactive intervention explores the intersection between new technologies, public space, and performance art. \cite{net_media_2023} The artist calls this phenomenon of relational architecture. \cite{lozano1999utterance} The intuitive approach to spontaneous real-time interaction utilised by \emph{Body Movies} played a pioneering role in shaping the design of urban media art in the subsequent two decades. The public square was transformed into a playground, offering the local people a new relationship with the city. 

In contrast to body interaction, \emph{Blinkenlights} by CCC was the first media facade work to use mobile devices to interact with the DIY media facade that the hackers made themselves. CCC transformed the Haus des Lehrers building in Berlin into a low-resolution (a window as pixel) but large-scale real-time interactive urban display. The audience could interact with the facade through emails and SMS. \cite{Blinkenlights2001} This project enabled the audience to spontaneously create verbal (through messages) and non-verbal (games and shapes) communication with the city.

The third example of interactive urban media art is \emph{LummoBlocks} by Lummo collective (Mar Canet, Carles Gutiérrez, Jordi Puig, and Javier Lloret). This art project applied a game strategy to engage the audience in play and create a new relationship with the neighbourhood. \emph{LummoBlocks} premiered in 2010 at MediaLab Prado urban screen in Madrid, funded from the city lighting budget to improve neighbourhood safety. Instead of playing alone on a small phone screen; the game experience was totally transformed: a nostalgic game Tetris in the size of a building, became a social phenomenon. One participant was controlling a shape and another location of a falling Tetris object. This way forces the audience to communicate. \emph{LummoBlocks} ran on the MediaLab Prado facade for nearly a year and became a meeting point for the locals to engage in play. \cite{Lummo_art_2010, canet2010LummoBlocks} A similar effect happened with the \emph{Pixel Killers} \footnote{https://var-mar.info/pixel-killers/} (2011) project by Varvara \& Mar at FACT Liverpool Disco window: interactive urban media art transformed the passing-by square into a social place for new experiences and encounters. \cite{Stubbs2012Disco}

In conclusion, interactive urban media art plays a pivotal role in reimagining public spaces, fostering novel urban experiences and enhancing communal interaction. Through innovative use of technology, these artworks invite audiences to engage directly with their surroundings, thus blurring the boundaries between art, technology, and daily life. The examples discussed demonstrate the variety and depth of interactive art's ability to transform urban environments into vibrant, active communities.

Compared to interactive art projects, participatory art offers no immediate response to the audience and resembles a ritual during a certain time. \cite{guljajevainteraction} While interaction creates the context for formally evoking real-time spontaneous reactions, engaging the audience often without verbal communication, participatory art generates activity for the audience with intention and organisation. Participatory art in the urban context often takes the form of an organised event with participants willingly contributing their creative outcomes to an artwork resulting from a collective effort, often with an underlying intention to promote a certain cause or carry out activism. Participatory art aims to offer a sense of community and authorship to the locals throughout its process. \cite{ellery_toward_2021}

To illustrate this, we chose to discuss Daniel Canogar’s series of works named \emph{Asalto} that took place in different cities starting in 2009 in Alcázar of Segovia, Spain. The artist placed a green screen horizontally and asked local people to crawl as if climbing an iconic building in their city. Later, Canogar reproduced the crawling people as climbing ones, and horizontal movement became vertical. \cite{canogar_asalto} For the audience, it was a unique experience to participate in the artwork and climb the walls of their city, which is an unusual thing to do. \emph{Asalto} can often be seen as interactive artwork, but since audience participation happens not in real-time and there is no feedback between the artwork and the participants, it is clearly a participatory art piece that offers to the audience a unique experience and belonging.

The second example is \emph{\#futurefridays} (2016) by Hannah Rosenfeld. This is a social art project that is concerned about the neighbourhood's future. Hence, the artists asked residents to imagine and draw possible future visions of their home area. The people’s visions were shared as an urban projection. In addition to participating in the drawing process, the event stimulated communication and discussion in the community. This project showcases public art with the primary participation of the community. Public participation is at the heart of this project, with opportunities for community members to be involved in the art project's creative process and the participants' creations to be used as artistic expression. According to Hannah Rosenfeld, their team was excited about the number and diversity of participants. \cite{rosenfeld_prototyping_2016} Also, the art event \emph{The Land is Dance} (2023) by artist Betty Carpick is a good example of community-oriented participatory urban media art.

\emph{Digital Calligraffiti} (2016) by Michael Ang and Hamza Abu Ayyash is a social art project that uses urban media to empower the participants. More specifically, during calligraffiti workshops, the professionals introduced young refugees to this medium and taught them how to express themselves through it. After the workshop, the participants performed live by projecting their messages in urban space. In the words of the artists: “This project format allows all social groups to freely share their messages and wishes. Using the art of calligraffiti to catalyse the communication between the unit and the whole, \emph{Digital Calligraffiti} aims at transforming the urban screens into a canvas of expression.” \cite{publicartlab2017digitalcalligraffiti}

In summary, this chapter delineates the distinct yet complementary roles of interaction and participation in urban media art. By examining these differences, we gain a deeper understanding of how these artworks engage communities, transform public spaces, and contribute to the cultural vibrancy of urban environments. The following chapter delves into a case study that effectively combines these two strategies, thereby enriching the process of place-making and audience engagement.
 
\section{We are the Clouds: Combining Audience Participation and Interaction}

\begin{figure*}[h]
\includegraphics[width=\textwidth]{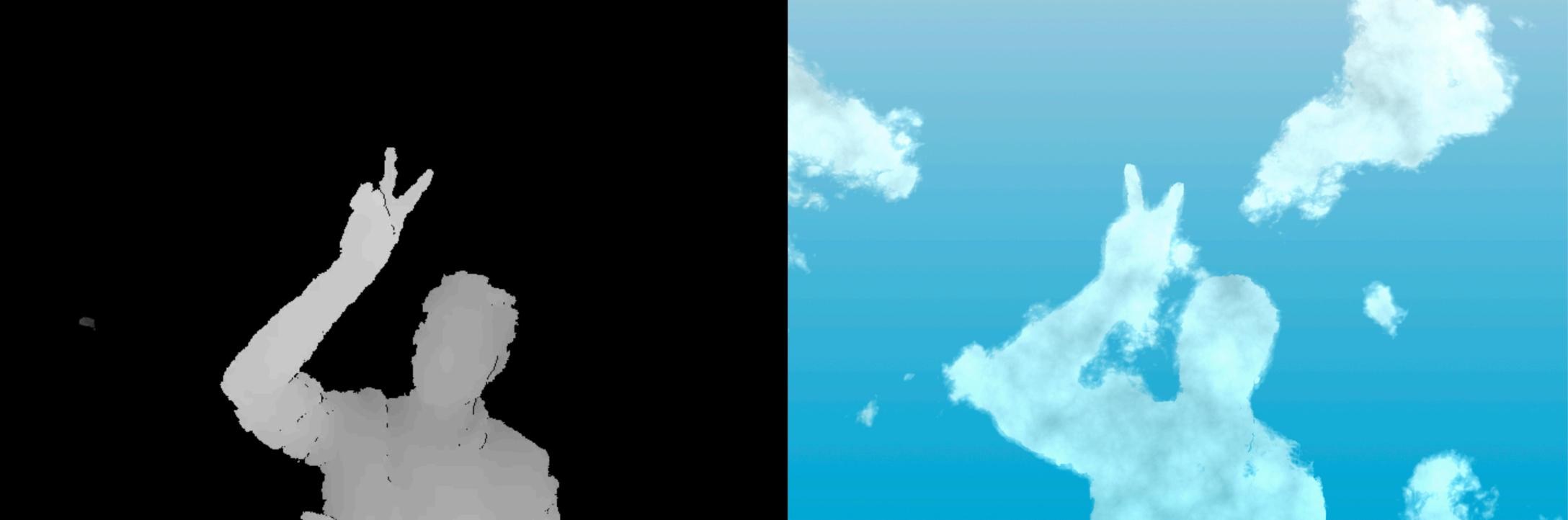}
\caption{Composition of different video stills from the depth-camera recordings. The participants using their creativity to make playful videos. \copyright Varvara \& Mar.}
\label{fig:composition}
\end{figure*}

\begin{figure*}[h]
\includegraphics[width=\textwidth]{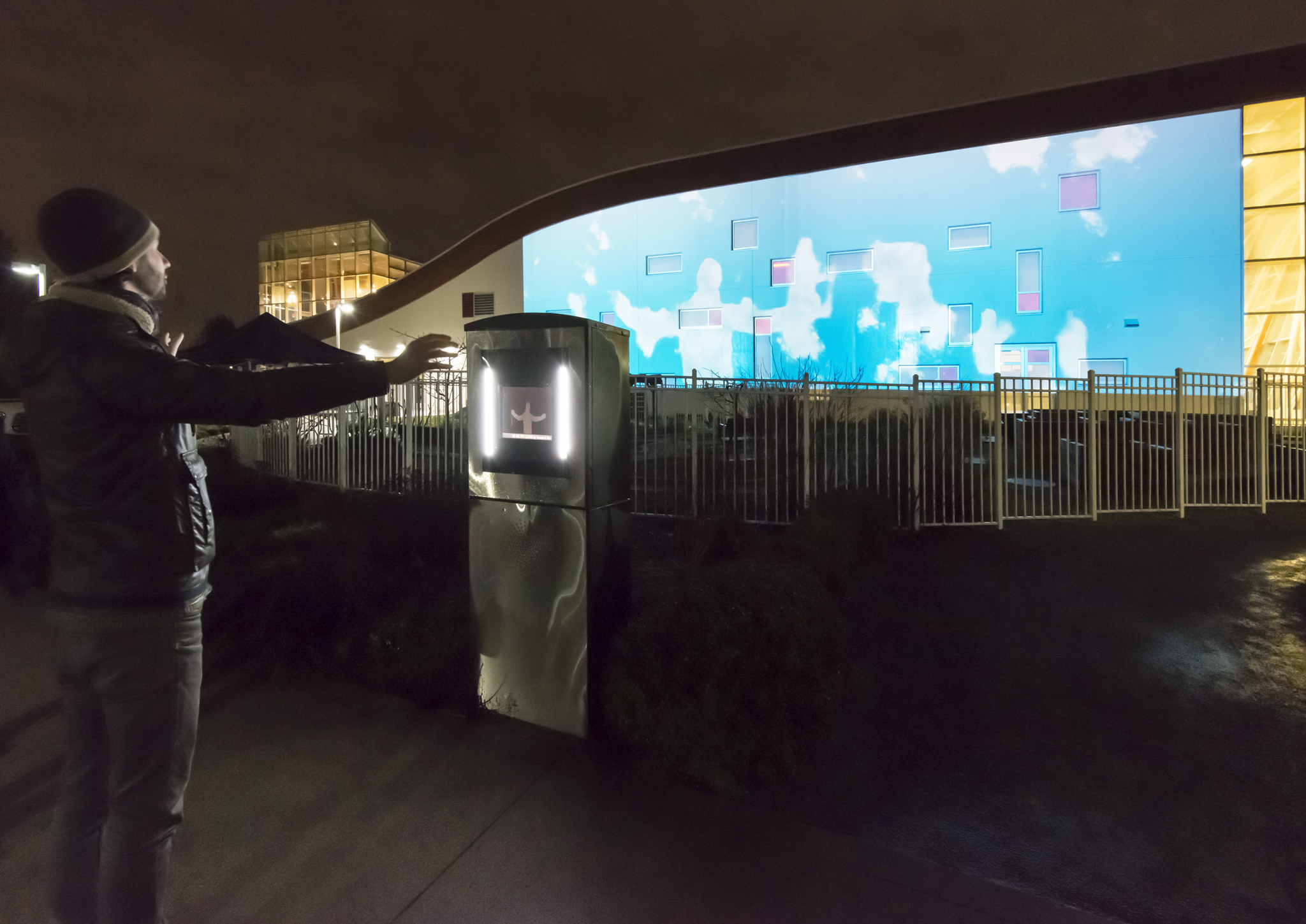}
\caption{The audience interacts with \emph{We Are The Clouds} urban media artwork at UrbanScreen.  \copyright Varvara \& Mar.}
\label{fig:interaction}
\end{figure*}

Moving from the theoretical framework to a specific example, our next focus is \emph{We Are The Clouds}. This artwork represents a fusion of interaction and participation within a single piece, shedding light on what can be achieved when these elements are applied together. By integrating both approaches, \emph{We Are The Clouds} offers a unique perspective on how combined interaction and participation can enrich the audience's experience, foster community engagement, and animate public spaces in meaningful ways. This case study will help us delve into the dynamics of such a combination, exploring its effects on both the artwork and its audience, and examining its contributions to the liveliness and social cohesion of urban settings.

In 2019-2020, The artist duo Varvara \& Mar were commissioned by Surrey Art Gallery to create new artwork for the digital facade at the Chuck Bailey Recreation Centre in Surrey, Vancouver. As mentioned before, UrbanScreen of Surrey Art Center was among the few art programs with an annual urban media art program. This venue showed and commissioned digital and interactive art from 2010 until 2022. \cite{surrey_urbanscreen} The curatorial concept and mission behind the digital canvas was an engagement with local ethnically vibrant communities and supporting artists with their novel concepts. \cite{artafterdark_2020}
Understanding Surrey's unique social fabric, the commissioned artist duo crafted \emph{We Are The Clouds} to not only reflect but also actively involve the local community (see Figure \ref{fig:general}). The artwork was conceptualised to integrate the audience in two fundamental ways: through participation in its creation and real-time interaction at the installation site. This dual approach allowed visitors to experience both the participatory and interactive elements simultaneously, fostering a deeper connection with the artwork.

\emph{We Are The Clouds} stands out as an interactive and participatory urban media artwork where the audience plays a pivotal role. Participants could transform their silhouettes into cloud shapes projected onto a large outdoor projection, simulating a dynamic skyscape at the UrbanScreen. Stacey Koosel writes the following about the piece: “To observe the clouds formed by previous visitors drift by is both playful and poignant, like reading through an animated guest book and trying to imagine the faces and stories behind the entries. Together, these visitors form a community of the often overlooked and underestimated art-viewing public. The act of deliberately creating clouds, which we would normally observe and interpret passively, is perhaps the most meaningful part of the piece.” \cite{koosel_softfluffypower_2020} Beyond its aesthetic appeal, the artwork also carried a critical layer of meaning. The metaphor of becoming a cloud paralleled the contemporary reality of digital existence, where significant aspects of our lives are stored in the virtual 'cloud', highlighting societal digitalization and vulnerability.

As mentioned before, the participatory part of the project consisted of a week-long workshop prior to the opening. The locals were invited to become a part of the art project by creating clouds with their own bodies that would drift over the digital sky of the UrbanScreen. The participants were encouraged to express what they would do as a cloud, resulting in diverse and imaginative performances captured by a Kinect depth camera. The workshop was ongoing for a week. People could drop in and out whenever they wished. This strategy enabled word to spread around and more people to attend. Some participants came back several times because they got new ideas for performing. For example, some came back with a friend and performed together in front of a depth camera; others used an object (see Figure  \ref{fig:participants_videos}). In terms of technical set-up in the workshop, the artists had a Kinect connected to a computer that was recording the participants' silhouettes. The people did not have a time limit for their moves, but the recommended length was about 30 sec. In total, 150 videos were recorded during the workshop, meaning Surrey people created as many cloud characters that would dance, play or walk on the urban screen.

The artists created C++ software using OpenFrameworks that transformed Kinect-recorded videos into clouds with shaders in real-time (see Figure \ref{fig:composition}). At a site where \emph{We Are the Clouds} was up on the urban screen was a dedicated interaction booth that contained another Kinect camera, a screen for the feedback of interaction and instruction purposes, and a computer.

In terms of the interaction, when one stands in front of the urban art installation, a Kinect camera picks up the silhouette and transforms it into a cloud in the digital sky in real-time (see Figure \ref{fig:interaction}). The participant’s body is a bigger cloud superimposed in the centre of the digital facade along with the pre-recorded participants, who are also converted into other smaller clouds that are passing by. In other words, all the clouds are different people creating a unique skyscape composition. The smaller clouds on the urban projection are the local people, who have contributed to their movements earlier. Hence, it is essential to note that the artwork brings together audience participation and real-time interaction on-site.

The participatory element of \emph{We Are the Clouds} is the pre-recorded videos of workshop participants, which in random order and position, appear and drift at different speeds over the digital sky as if they were clouds. \emph{We Are The Clouds} is a generative and interactive art piece, which means that the content is always different. It is unique each time, like the real sky, mirroring the participation of the local people. 

To sum up, \emph{We Are The Clouds} exemplifies the successful integration of audience participation and interaction in urban media art. By inviting community members to be both creators and interactors, the artwork transcends traditional boundaries, fostering a sense of ownership, engagement, and creativity. This project not only enlivens the urban landscape but also contributes to the communal identity and placemaking, demonstrating the potential of combining participatory and interactive elements in urban media art.

\section{Synergising Participation and Interaction in Urban Media Art: A Dynamic Engagement}

Our exploration of participatory and interactive art forms within the urban media context reveals profound insights into how these approaches foster community engagement and transform public spaces. The participatory art examples, as discussed earlier, underscore the importance of involving the audience in the creative process. As Claire Bishop noted, this involvement is pivotal for activation, authorship, and fostering a sense of community. \cite{bishop_participation_2010} Participatory art typically culminates in video-based installations, offering audiences the opportunity to reflect and recognise their collective efforts.

Interactive urban media art, contrastingly, thrives on real-time engagement, creating a dynamic and responsive environment. As Lozano-Hemmer poignantly remarks: “My pieces do not exist unless someone dedicates some time to them.” \cite{lozano2001metaphors} Audience interaction is a very engaging element in the case of urban media art, as the aforementioned \emph{LummoBlocks} or \emph{BlinkenLight} illustrated. Hence, interactive urban media artworks can contribute to community building and reclaiming public space, but there is also a danger for interactive works to become a banal spectacle because interaction has a strong connotation of attraction. However, it also treads a fine line between engaging spectacle and banal attraction, emphasising the critical role of thoughtful curation and sustained art programming in maintaining the depth and relevance of these installations.

In comparing participatory and interactive approaches, we can consider Daniel Canogar's \emph{Asalto} and \emph{We Are The Clouds} by Varvara \& Mar. Both projects involve the audience in the art-making process, yet \emph{Asalto} remains purely participatory, with the final outcome observed rather than influenced by the viewers. This contrast highlights a key characteristic of participatory art – its potential for immersive experience is often limited to the period of creation. In stark contrast, \emph{We Are The Clouds} combines participation with interaction, allowing ongoing engagement with the artwork, thereby expanding the scope and impact of audience involvement. According to our study, this unique fusion of participation and interaction is rarely seen in urban media art. The closest urban media artwork demonstrating a similar strategy is \emph{Vectorial Elevation} (1999) by Rafael Lozano-Hemmer. Although light beams were not controlled in real-time by the audience, the software simulation was interactive. The participants designed their light animations on the web, which were later performed in urban space. \cite{lozano_hemmer_vectorial_1999} This project represents an early attempt to blend two audience engagement methods.

Our study highlights that urban media artworks combining participation and interaction are exceptional yet highly effective in engaging local communities. They provide a comprehensive model for community engagement, offering both a sense of ownership in the creative process and a dynamic interaction with the final artwork. While the prevalence of such integrative approaches remains limited, their potential for enriching the social fabric and enhancing the experiential quality of urban spaces is undeniable.

\section{Conclusions}

This article has explored the nuanced differences and interplay between audience interaction and participation in urban media artworks, specifically through the lens of the case study \emph{We Are The Clouds} Our examination has highlighted the distinct yet complementary roles of interaction and participation in enhancing the community's engagement with public space, fostering a sense of belonging, and contributing to the cultural identity of urban environments.

Audience interaction, characterised by its immediacy and real-time engagement, transforms urban spaces into dynamic arenas for spontaneous creativity and social interaction. It invites the public to participate actively in the artwork, offering a unique and immersive experience that enriches the urban landscape. In contrast, participatory art emphasises a more deliberate and organised involvement of the community, often culminating in a collective artistic outcome that reflects a shared experience or message.

\emph{We Are The Clouds} is an example of how these two methods can be effectively combined in urban media art. By integrating both real-time audience interaction and pre-event participatory elements, the artwork offers a multifaceted experience that is both inclusive and dynamic. It allows for the expression of individual creativity and the formation of a collective narrative, encapsulating the diversity and dynamism of the local community.

Our research suggests that such a dual approach is not commonly utilised in urban media art, though it holds significant potential for deeper community engagement. When both interaction and participation are employed, they create a more holistic experience, encompassing both the immediate thrill of interaction and the reflective depth of participation. This combination can lead to a richer, more meaningful connection between the artwork, its audience, and the urban context.

In conclusion, the exploration of \emph{We Are The Clouds} reaffirms the importance of integrating both interaction and participation in urban media art. Such an approach not only enhances the aesthetic and experiential qualities of the artwork but also plays a crucial role in redefining public spaces as vibrant, inclusive, and culturally significant areas. It encourages a reimagining of urban environments as canvases for creative expression and communal storytelling, fostering a stronger sense of community and place-making.

\section{Acknowledgments}
 \emph{We Are The Clouds} was commissioned by Surrey Art Gallery. Mar Canet Sola is supported as a CUDAN research fellow and ERA Chair for Cultural Data Analytics, funded through the European Union’s Horizon 2020 research and innovation program (Grant No. 810961).

\bibliographystyle{isea}
\bibliography{bibliography}

\section{Author(s) Biography(ies)}
Dr Varvara Guljajeva is an Assistant Professor in Computational Media and Arts at the Hong Kong University of Science and Technology (Guangzhou). Previously, she held positions at the Estonian Academy of Arts, Elisava Design School in Barcelona, and and the University of Art and Design Linz. Her PhD thesis “From Interaction to Post-Participation: The Disappearing Role of the Active Participant,” was selected as the highest-ranking abstracts by Leonardo Labs in 2020. Prof Guljajeva has published her research in numerous academic venues, such as Siggraph, ISEA, IEEE VISAP and more. In 2021, she co-edited the book ‘Meaning of Creativity in the AI-Age’. Varvara works with Mar Canet as an artist, forming an artist duo, Varvara \& Mar. 

Mar Canet Sola is a PhD candidate and research fellow at Cudan research group in BFM Tallinn University. He has a master’s degree from Interface Cultures at the University of Art and Design Linz and two degrees in art and design from ESDI in Barcelona and in computer game development from University Central Lancashire in the UK. Mar has presented his research in prominent academic conferences, like Siggraph, ISEA, TEI, and more. As an artist, he works together with Varvara Guljajeva, forming an artist duo Varvara \& Mar. 

\end{document}